\documentclass[10pt, conference]{IEEEtran}
\IEEEoverridecommandlockouts
\usepackage{graphicx}
\usepackage{color, verbatim}
\usepackage{cite}
\usepackage{latexsym}
\usepackage{verbatim}
\usepackage{booktabs,tabularx}
\usepackage{epsfig}

\usepackage[cmex10]{amsmath}
\usepackage{amsmath}
\DeclareMathOperator*{\argmax}{argmax}
\usepackage[cmex10]{amsmath}
\usepackage{amssymb}
\usepackage{mathtools}
\usepackage [english]{babel}
\usepackage [autostyle, english = american]{csquotes}
\newcommand{\SNR}{\textrm{SNR}}

\MakeOuterQuote{"}
\usepackage{mathtools}
\usepackage{geometry}
\begin{document}
\pagestyle{empty}  

\title{Beamformed Energy Detection in the Presence of an Interferer for Cognitive mmWave Network}

% \author{M.~Madhuri~Latha$^1$, Sai Krishna~Charan~Dara$^1$, Sachin~Chaudhari$^1$, Neeraj Varshney$^2$ \\ 
% $^1$Signal Processing and Communication Research Center, IIIT Hyderabad, India\\
% $^2$Wireless Networks Division, NIST, Gaithersburg, USA\\
% %International Institute of Information Technology-Hyderabad, India \\
%  madhuri.latha@research.iiit.ac.in, sai.krishna@students.iiit.ac.in, sachin.chaudhari@iiit.ac.in,
% neeraj.varshney@nist.gov
% }
\author{
\IEEEauthorblockN{M.~Madhuri~Latha$^1$, Sai Krishna~Charan~Dara$^1$, Sachin~Chaudhari$^1$, Neeraj Varshney$^2$}
\IEEEauthorblockA{\textit{$^{1}$Signal Processing and Communication Research Center, IIIT Hyderabad, India}\\
\textit{$^2$Wireless Networks Division, NIST, Gaithersburg, USA}\\
Emails: madhuri.latha@research.iiit.ac.in, sai.krishna@students.iiit.ac.in, sachin@iiit.ac.in,
neeraj.varshney@nist.gov}
}
% \author{M.~Madhuri~Latha, Sai Krishna~Charan~Dara, Sachin~Chaudhari\\ 
% Signal Processing and Communication Research Center,\\
% International Institute of Information Technology-Hyderabad, India \\
% Email: \{madhuri.latha@research.,sai.krishna@students., sachin.chaudhari@\}iiit.ac.in
\maketitle

\begin{abstract}
In this paper, we propose beamformed energy detection (BFED) spectrum sensing schemes for a single secondary user (SU) or a cognitive radio to detect a primary user (PU) transmission in the presence of an interferer. 
In the millimeter wave (mmWave) band, due to high attenuation, there are fewer multipaths, and the channel is sparse, giving rise to fewer directions of arrivals (DoAs). Sensing in only these paths instead of blind energy detection can reap significant benefits. An analog beamforming weight vector is designed such that the beamforming gain in the true DoAs of the PU signal is maximized while minimizing interference from the interferer. To demonstrate the bound on the system performance, the proposed sensing scheme is designed under the knowledge of full channel state information (CSI) at the SU for the PU-SU and Interferer-SU channels. However, as the CSI  may not be available at the SU, another BFED sensing scheme is proposed, which only utilizes the estimate the DoAs. To model the estimates of DoAs, perturbations are added to the true DoAs. The distribution of the test statistic for BFED with full CSI schemes is derived under the null hypothesis so that the threshold of the Neyman-Pearson detector can be found analytically. The performance of both schemes is also compared with the traditional energy detector for multi-antenna systems.

%Using these DoAs, an analog beamforming weight vector is designed such that beamforming gain in the estimated DoAs of the PU signal and nulling interference from interferer is maximized. First, a BFED scheme is designed using estimates of the DoAs. The performance of this proposed scheme is also compared with BFED, which is designed with full CSI.  It is shown through simulations that the proposed BFED method has significant detection performance compared to a simple energy detector even without knowing channels statistics between PU and interferer to the secondary user (SU). Also, the deterioration in the detection performance using analog beamforming does not deteriorate significantly even if CSI corresponding to the PU-SU and Interferer-SU channels is not available.
\end{abstract}


%%%%%%%%%%%%%%%%%%%%%%%%%%%%%%%%
\section{Introduction}  
%%%%%%%%%%%%%%%%%%%%%%%%%%%%%%%%
%----Why mmWave systems?
With an exponential increase in the number of wireless devices, services, and data usage, the availability of high-quality spectrum has become a bottleneck for the next-generation wireless system. To address these requirements, millimeter wave (mmWave) bands from 30 GHz to 300 GHz with huge bandwidths have been proposed to be a key enabler for 5G \cite{Niu2015, Kutty2016}. 
%------Cognitive Radio in the presence of mmWave
It is most likely that a mmWave network including that of 5G is going to be a heterogeneous network \cite{Hosseini2016, Gupta2016, Tsinos2020}. Also, spectrum sharing is proposed in \cite{Gupta2016} for different operators in the 5G network on the same frequency. There may also be existing incumbents in mmWave bands such as satellite communications, research, military, and unlicensed operations \cite{Hosseini2016, Gupta2016, Tsinos2020}. Interference coming from heterogeneous mmWave networks sharing the same band can have a negative impact on the achieved throughput and reliability due to the interference even with narrow beams \cite{Gupta2016, Park2018, Tsinos2020}. Cognitive radio (CR) is a potential technology that can address the problem of interference among the coexisting heterogeneous mmWave wireless systems and improve their performance \cite{ mmWaveCR, Tsinos2020}.
\begin{comment}
{
Cognitive radio can provide solutions to reduce the interference among the coexisting heterogeneous wireless systems and improve their performance \cite{Thesis, Gao2012}.  
}
\end{comment}

%------ Spectrum Sensing in the presence of interferer 
Spectrum sensing is an important CR technology that provides spectrum awareness and managing interference among heterogeneous mmWave networks in 5G. Several types of spectrum sensing techniques have been proposed in cognitive radio paradigm: energy, feature, and matched filter-based \cite{Thesis}. However, energy detection (ED) has been widely adopted under fading channel due to its simplicity since the primary user (PU) information is not required. 
%%%%%%%%%%%%%%%%%%%%%%%
\begin{comment}
Also, ED is an optimal scheme for detecting a random signal in additive white Gaussian noise (AWGN) under the assumption of noise statistics \cite{KayDetTheory}. 
\end{comment}
%%%%%%%%%%%%%%%%%%%%
%-------Why Beamforming for sensing?
Traditionally, most of the work on spectrum sensing, including ED, has assumed using one or more omni-directional antennas \cite{Thesis}. However, beamforming based sensing can improve the detection performance over omni-directional sensing \cite{NCC}. The use of beamforming for data transmission is imperative at mmWave frequencies, where the signal undergoes severe propagation loss and travel in a highly directional manner leading to fewer multipaths \cite{Niu2015}. Given that a massive number of antennas can be packed in a small form factor at mmWave frequencies, beamforming can be extremely fruitful for spectrum sensing as well. 

The existing literature has limited works that are focused on receiver beamforming for sensing \cite{ BF_Sensing_1, BF_Sensing_2, NCC}. An eigenvalue-based spectrum sensing algorithm is proposed in \cite{BF_Sensing_1} using a beamformed received signal. In \cite{BF_Sensing_2}, the angular domain is divided into sectors, and these sectors are then sensed serially using beamforming. Most of the spectrum sensing schemes in the literature, including ED, only assume additive noise at the receiver and ignore any interference caused by a non-cooperating secondary user or unregulated transmission. The presence of interfering node can significantly degrade the detection performance. This issue is even more aggravated in mmWave networks, which are heterogeneous, as explained before. In this context, the works in \cite{Makarfi2012, Lagunas2014} address spectrum sensing of a PU in the presence of an interferer . In \cite{Makarfi2012}, the performance of a sensing node is analyzed in a multi-user environment with the presence of interference from unlicensed users of a non-cooperating secondary network. In \cite{Lagunas2014}, several compressive spectrum sensing schemes are compared for detecting PU frequencies in the presence of interference from low-regulated transmissions from unlicensed systems. However, both the sensing algorithms have been suggested for traditional cognitive networks and not for the cognitive mmWave networks. Also, no receiver beamforming is assumed in both sensing schemes. Beamformed energy detection (BFED) in the cognitive mmWave network was recently proposed in \cite{ANTS} to improve the sensing performance over clustered Rician fading channel. However, the impact of interferer is not considered while analyzing the sensing performance. 

%------our contribution
In this paper, we propose BFED\footnote{Although the results of this paper may be extended to other sensing schemes such as maximum eigenvalue detection, maximum to minimum eigenvalue detection \cite{spectrum_sensing_survey}, we limit ourselves to the ED for convenience.} spectrum sensing schemes in the presence of an interferer for cognitive mmWave networks.  The multipath channel considered in this paper is the extended Saleh-Valenzuela channel model \cite{channel}, which is suitable to model mmWave channels as most of the traditional fading distributions do not work for the mmWave channel. The test statistic at the receiver is the energy of the signal received using analog beamforming. For this BFED test statistic, we employ the Neyman-Pearson detector, which maximizes the probability of detection for a given false alarm. To improve the detection performance, it is important to choose optimal beamforming weights such that the probability of detection is maximized. However, such optimization will be non-linear and non-trivial in terms of the weight vector. This is true even for optimizing SINR. Therefore, we attempt to solve a multi-objective function that tries to maximize the beamforming gain along the direction of arrivals (DoAs) corresponding to the PU while creating a null along the DoAs of the interference signal. Note that the proposed sensing scheme is designed under the knowledge of full CSI at the SU for the PU-SU and Interferer-SU channels to demonstrate the bound on the system performance. As the channel information may not be available at the SU, a second sensing scheme is also proposed that does not require CSI. In this case, the focus is only on estimating the DoAs as estimating full CSI may be difficult or infeasible. In this paper, we abstract out the DoA estimation schemes by adding an error or perturbation to the actual angles, which is modeled as Gaussian distributed with mean zero and variance linked to the Cramer Rao Lower Bound (CRLB) for the problem \cite{CRLB}. For both the proposed BFED schemes, the distribution of the test statistic is derived under the null hypothesis under the assumption of knowing noise and interference powers so that the threshold for a Neyman-Pearson detector can be found analytically. The performance of both the proposed BFED schemes is also compared to a simple ED.

The rest of the paper is structured as follows. Section \ref{Sec:SystemModel} describes the system model. Section \ref{Sec:BFEF-FCSI} and \ref{Sec:BFEF-PDoA} present the proposed BFED with perfect CSI and estimated or perturbed DoAs. Section \ref{Sec:Results} presents simulation results while  concluding remarks are given in Section \ref{Sec:Conclusion}.

\section{System Model} \label{Sec:SystemModel}
Fig. 1 shows the schematic diagram of the CR based mmWave network considering a PU, a SU, and an interferer in the scenario. The presence of a PU is detected using SU in the presence of an interferer. It is assumed that the SU has $M$ antennas arranged as uniform linear array (ULA) with the separation between the antennas as $d=\lambda/2$, where $\lambda$ is the wavelength of the transmitted PU signal. Both the interferer and PU are assumed to have a single antenna for simplicity. This assumption is valid for simple IoT nodes with low-power consumption due to one RF chain.    
\begin{figure*}[t!]
	\centering
	\includegraphics[height= 2.4in,width=4in]{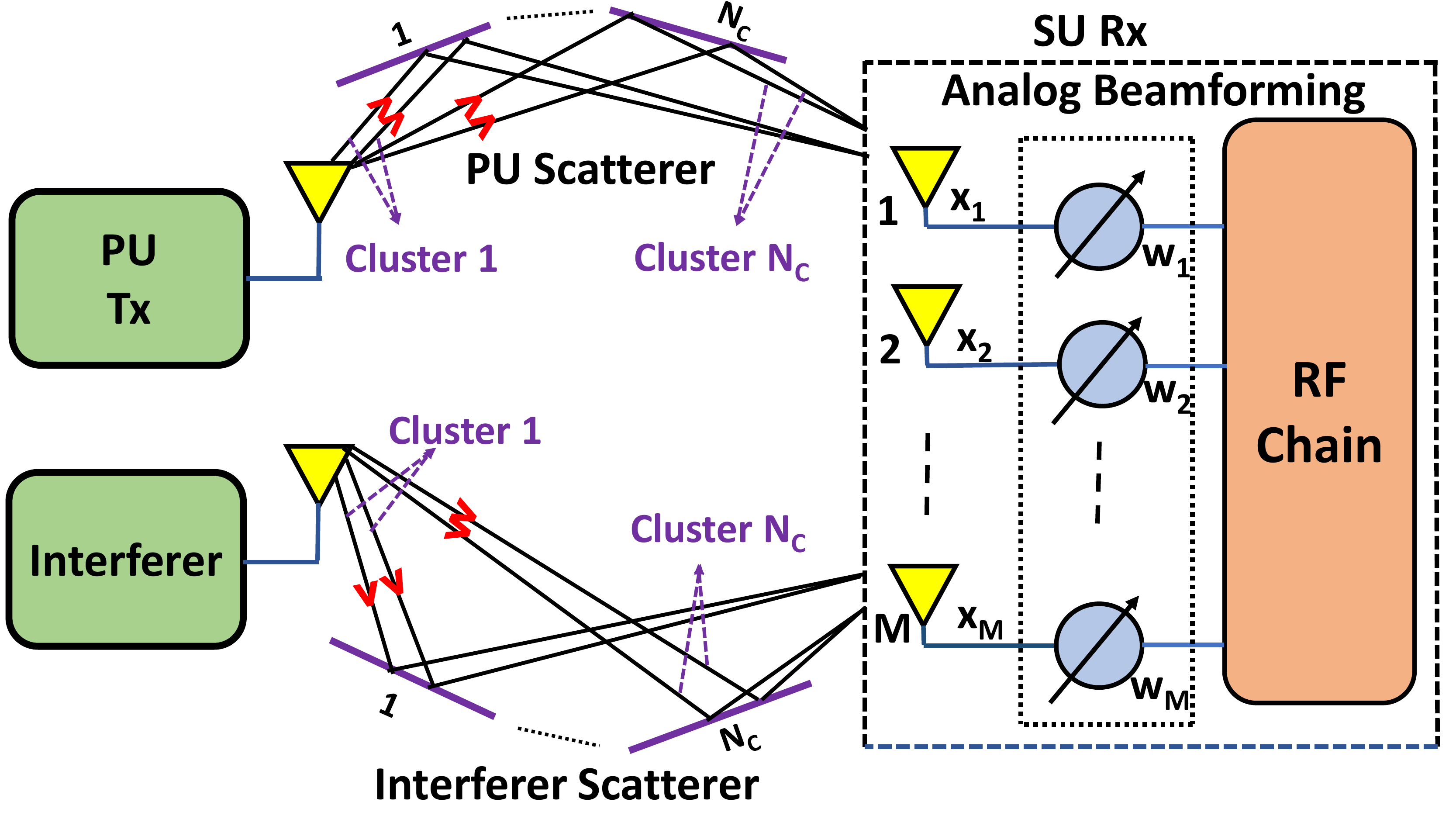}
% 	\vspace{-0.3cm}
	 \caption{Schematic diagram of the CR based mmWave network considering the presence of a PU, a SU and an interferer.}
% 	 	\vspace{-0.3cm}
    \label{fig:case 1}
\end{figure*}
In spectrum sensing, PU signal is detected by deciding between the two hypotheses $\mathcal{H}_0$ and $\mathcal{H}_1$, where $\mathcal{H}_0$ denote the absence of PU signal and $\mathcal{H}_1$ denote the  presence of PU signal. Under the two hypotheses, the received signal $\mathbf{x}[n]\in \mathbb{C}^ {M \times 1}$ for $n=1, 2, \ldots, N$, where $N$ is the number of snap shots, is given by
\begin{equation} \label{eq:xn}
\begin{split}
\mathcal{H}_0 &:\mathbf{x}[n] = \mathbf{h}_i[n]s_i[n]+\mathbf{v}[n], \\
\mathcal{H}_1 &: \mathbf{x}[n] = \mathbf{h}_p[n]s_p[n] + \mathbf{h}_i[n]s_i[n] + \mathbf{v}[n],
\end{split}
\end{equation}
where $s_p[n]$ and $s_i[n]$ represent the PU and the interferer signals, respectively. The quantity $\mathbf{v}[n]  \in \mathbb{C}^ {M \times 1}$ denotes the AWGN noise vector, which comprises of entries $v_m[n], m=1,2,\cdots,M$, where $v_m[n]$ is the AWGN noise sample at $m$th antenna. Note that each of the terms follows complex Gaussian distribution as  
$s_p[n]\sim\mathcal{CN}(0, \sigma^2_{sp})$,  $s_i[n]\sim\mathcal{CN}(0, \sigma^2_{si})$ and  $v_m[n]\sim\mathcal{CN}(0, \sigma^2_{v})$ respectively. 
%The observations at each of the $M$ antennas are independent of each other under ${H}_0$ hypotheses.
The vector $\mathbf{h}$ $\in$ $\mathbb{C}^ {M \times 1}$ (referring to either $\mathbf{h}_i$ or $\mathbf{h}_p$), is assumed to be the sum of signals from $N_{c}$ clusters each of which have $N_{r}$ rays.  In this work, we utilize the extended Saleh-Valenzuela mmWave channel model \cite{CRLB}. Under the model, the narrow-band channel  $\mathbf{h}$ can be described as \cite{channel}
\begin{equation}
 \mathbf{h}=\Gamma\sum_{c=1}^{N_c}\sum_{r=1}^{N_r}\alpha_{cr}\mathbf{a}_r(\theta_{cr})),
 \end{equation}
 %\vspace{-2mm}
 where
 \begin{itemize}
 \item{$\alpha_{cr}$ =} complex fading coefficient of the $r$th ray in the $c$th scattering cluster,
 \item{$\theta_{cr}$ =} azimuth DoA of the $r$th ray in the $c$th scattering cluster at the receiver,
 \item{$\Gamma$=}$\sqrt\frac{M}{N_{c}N_{r}}$ is a normalization constant. 
 \end{itemize}
Here, the complex fading coefficient $\alpha_{cr}$ follows complex Gaussian distribution as $\alpha_{cr}\sim\mathcal{CN}(0,1)$. The mean DoAs of the different clusters at the receiver follow a uniform distribution, and the DoAs of the rays within an individual cluster are distributed according to the Laplacian distribution function. The antenna array response vector at the receiver is given by 
%-------------------------
\begin{align}
 \mathbf{a}_{r}(\theta_{cr})=\frac{1}{\sqrt{M}}\big[1,&e^{-jkd\sin(\theta_{cr})},...,\nonumber\\
 &e^{-(M-1)jkd\sin(\theta_{cr})}\big]^T.
\end{align}
  %------------------------
 Under $\mathcal{H}_0$, the received signal at $m$th antenna is given as
 \begin{equation}
    {x}_m[n] = {h}_{im}[n]s_i[n]+{v_m}[n].
 \end{equation}
 
 \begin{comment}
 where 
 \begin{equation*}
 {h}_{im}[n]=\sqrt\frac{M}{N_cN_r}\sum_{c=1}^{N_c}\sum_{l=1}^{N_r}\alpha_{cl}e^{j(m-1)kd\sin(\theta_{cl})},
\end{equation*}
\end{comment}
The conditional mean and variance of the received signal $x_m[n]$ conditioned on the two hypotheses can be derived as
%----------------------------------
\begin{align}
 E\{x_m[n] \, | \, \mathcal{H}_0\} & =  0, \\ 
 E\{x_m[n] \, | \, \mathcal{H}_1\} & =  0,\\
 E\{\textrm{var}[x_m[n]] \, | \, \mathcal{H}_0\} 
 & = E\{|h_{im}[n]|^2\}\sigma^2_{si}+\sigma^2_{v} \nonumber\\
 & = \sigma^2_{si}+\sigma^2_{v},\\ 
 E\{\textrm{var}[x_m[n]] \, | \, \mathcal{H}_1\} & = E\{|h_{pm}[n]|^2\}\sigma^2_{sp}\nonumber\\  &\hspace{0.35cm}+E\{|h_{im}[n]|^2\}\sigma^2_{si}+\sigma^2_{v} \nonumber\\
 &= \sigma_{sp}^2+ \sigma_{si}^2+\sigma_{v}^2.
\end{align}
 %------------------------------------------  
 Therefore, the distributions of $x_m[n]$ under the two hypotheses are given by
\begin{equation} \label{dist_x_n_H0_H1}
\begin{split}
\mathcal{H}_0 &: x_m[n]\sim\mathcal{CN}\left(0, \sigma^2_{si}+\sigma^2_{v}\right),\\
\mathcal{H}_1 &: x_m[n]\sim\mathcal{CN}\left(0, \sigma_{sp}^2+ \sigma_{si}^2+\sigma_{v}^2\right).
\end{split}
\end{equation}
%-----------------------------------------------%
%%%%%%%%%%%%%%%%%%%%%%%%%%%%

\section{BFED Spectrum Sensing Scheme with Full CSI}\label{Sec:BFEF-FCSI}
%-------------------------------------------------------------
In this section, we assume the channels between PU $\rightarrow$ SU and interferer $\rightarrow$ SU are perfectly known at the SU to demonstrate the bound on the system performance.  Note that the receive beamforming is achieved by multiplying the received signal with a  analog beamforming weight vector $\mathbf{w}$. While designing the weight vector $\mathbf{w}$, two factors are considered:

(i) minimize the interference from the interferer and 

(ii) maximize the beamforming gain in the PU direction. 
\\
First we consider designing the weight vector to minimize the interference. To do this, the interference matrix $\mathbf{H}_i$ is constructed using the interferer $\rightarrow$ SU channel consisting of array steering vectors corresponding to all DoAs of interferer and their path gains so that
\begin{equation}
      \mathbf{H}_i=\big[\alpha_{i1}\mathbf{a}(\theta_{i1}), \ \alpha_{i2}\mathbf{a}(\theta_{i2}),\ \cdots,\ \alpha_{iK}\mathbf{a}(\theta_{iK}) \big]^H,
\end{equation}
where $K=N_cN_r$. The beamforming weight vector $\mathbf{w}$ has to be chosen such that it lies in the null space of interference matrix. Apply singular value decomposition (SVD) to obtain the null space of $\mathbf{H}_i$. It is given as
\begin{equation}
      \mathbf{H}_i=\mathbf{U}\mathbf{\Sigma}[\mathbf{V}_{1}\mathbf{V}_{0}]^H,
\end{equation}
where $\mathbf{V}_1$ holds the first $K$ right singular vectors and  $\mathbf{V}_0$ holds the last $M-K$ right singular vectors of $\mathbf{H}_i$. Note that $M > K$ is a valid assumption as mmWave channels have fewer multipaths because of limited scattering. Now, using the last $M-K$ right singular vectors,  the projection vector $\mathbf{w}_h$ from $\mathbf{w}$ to the null space of $\mathbf{H}_i$ can be obtained as \cite{Jiang2019}
\begin{equation}
\mathbf{w}_h = \mathbf{V}_{0}{{\mathbf{V}_{0}}^H}\mathbf{w}.
\end{equation}
Here, for simplicity, we maximize the square norm of $\mathbf{w}_h$ which is given by
\begin{equation}\label{eq:Interference}
   ||{\mathbf{w}_h}||^2=\mathbf{w}^H\mathbf{V}_{0}{\mathbf{{V}_{0}}^H}\mathbf{w}.
\end{equation}

Now considering the second factor for designing the weight vector so that beamforming gain in the PU direction is maximized. Tuning to the DoA corresponding to the PU gives beamforming gain which is given in \cite{Jiang2019} by
\begin{equation}\label{eq:BG}
B_G=\mathbf{w}^H \mathbf{H}_p \mathbf{H}_p^H\mathbf{w},
\end{equation}
% where $\mathbf{H}_p=[\ \alpha_{p1}\mathbf{a(\theta}_{p1}), \ \alpha_{p2}\mathbf{a(\theta}_{p2}),\ \ldots\ \alpha_{pK}\mathbf{a(\theta}_{pK}) \ ]^H $ is the matrix consisting of channel vectors, consisting of steering vectors corresponding to the DoAs from PU and their channel gains. 
where $\mathbf{H}_{p}=\left[\alpha_{p 1} \mathbf{a}\left(\theta_{p 1}\right), \alpha_{p 2} \mathbf{a}\left(\theta_{p 2}\right), \cdots, \alpha_{p K} \mathbf{a}\left(\theta_{p K}\right)\right]^{H}$
matrix consists of steering vectors corresponding to the DoAs from PU and their channel gains.

Next, we need to combine these two objective functions (\ref{eq:Interference}) and (\ref{eq:BG}) together so that the analog beamforming weight vector can be designed to maximize the beamforming gain while minimizing the interference from the interferer by projecting onto the null space. It is a multi objective optimization problem with a constraint that the magnitude of each element in the weight vector should be constant, which is a non-convex constraint. It is worth noting that the Pareto optimality is usually used to describe solution of a multi-objective problem. Here, we have used weighted-sum method which can provide a sufficient condition for Pareto optimality assuming all the weights are positive and the summation of the weights is equal to one \cite{Jiang2019}. The multi-objective problem can now be written as
% \begin{equation}
% \begin{split}
% \mathbf{w}_{\opt} &= \, \argmax_\mathbf{w}       \Big\{\lambda \, \mathbf{w}^H\mathbf{V}_{0}{\mathbf{V}_{0}^H} \mathbf{w}
% \\ &\qquad \qquad + (1-\lambda)\,\mathbf{w}^H \mathbf{H}_p \mathbf{H}_p^H\mathbf{w}\Big\}, \\
% & \quad \textrm{s.t.} \quad\mathbf{w} \in \mathcal{F},
% \end{split}
% \end{equation}
%Please look into this equation
\begin{equation}
\begin{aligned}
\mathbf{w}=\underset{\mathbf{w}}{\operatorname{argmax}} &\left\{\lambda \mathbf{w}^{H} \mathbf{V}_{0} \mathbf{V}_{0}^{H} \mathbf{w}\right.\\
&\left.+(1-\lambda) \mathbf{w}^{H} \mathbf{H}_{p} \mathbf{H}_{p}^{H} \mathbf{w}\right\} \\
\quad  \text { s.t. } \mathbf{w} \in \mathcal{F}
\end{aligned}
\end{equation}
where $0<\lambda<1$ and $\mathcal{F}$ is a set of all vectors, where each element in the vector has constant magnitude. Note that solving the above problem is difficult because of non-convex constraint. However, after transforming the problem into semi definite programming through algebraic transformation, this problem can be simplified as
\begin{equation}
\begin{split}
 \mathbf{W}_{1} &=  \argmax_{\mathbf{W}} \, \Big\{\textrm{Tr}\big( (\lambda \, \mathbf{V}_{0}{\mathbf{V}_{0}^H} \\
& \qquad \qquad + (1-\lambda) \, \mathbf{H}_p \mathbf{H}_p^H ) \mathbf{W} \big) \Big\},\\
& \textrm{s.t.} \quad
  [\mathbf{W}]_{kk} =\frac{1}{M} \quad \forall k=1, \cdots, M, \\
& \quad \qquad \mathbf{W}\geq 0, \quad \textrm{Rank}(\mathbf{W})=1,
 \end{split}
\end{equation}
where $\mathbf{W}=\mathbf{w}\mathbf{w}^H$ and $[\mathbf{W}]_{kk}$ is the $k$th diagonal element of $\mathbf{W}$. Here, $\textrm{Tr}(.)$ refers to trace of a matrix. However, the rank one constraint in the above optimization problem is hard to solve. To deal with this, the optimization problem is modified by dropping rank one constraint \cite{Jiang2019}. It is given as
\begin{equation}
\begin{split}
 \mathbf{W}_{2} &=  \argmax_{\mathbf{W}} \, \Big\{\textrm{Tr}\big((\lambda\,\mathbf{V}_{0}{\mathbf{V}_{0}^H} \\
& \qquad \qquad + (1-\lambda) \, \mathbf{H}_p \mathbf{H}_p^H \big) \mathbf{W}) \Big \},  \\
& \textrm{s.t.} 
 \quad [\mathbf{W}]_{kk}=\frac{1}{M} \quad \forall k=1, \cdots, M, \\ 
& \quad \qquad \mathbf{W}\geq 0.    
 \end{split}
\end{equation}
Let $\bf{w}_2$ be the first column of $\bf{W}_2$. Now the analog beamforming weight  vector $\mathbf{w}_*$ is taken to be scaled version of $\bf{w}_2$, given by
\begin{equation}
    {\bf{w}}_{*} =  \sqrt{M}\bf{w}_2.
\end{equation}
Now, each element in the weight vector calculated from the optimization problem may not always be of constant magnitude $\frac{1}{\sqrt{M}}$ as  rank one constraint is ignored. To deal with this problem, we choose a sub-optimal solution in which each element in the weight vector is   divided by its magnitude. The weight vector after normalization is
\begin{equation}{\label{eq:mag}}
    {w}_{CSI}[k]=\frac{w_*[k]}{\sqrt{M}|w_*[k]|},\quad \forall k=1,\cdots, M.
\end{equation}

Next, the received signal is multiplied by the weight vector so that the beamformed received signal is given by
\begin{equation}
    z[n] = \mathbf{w}_{CSI}^H \mathbf{x}[n],
\end{equation}
while the energy $E_{CSI}$ can be calculated from beamformed received signal by 
\begin{equation}
E_{CSI} = \sum_{n=1}^{N}|z[n]|^2.
\end{equation}
The distribution of $E_{CSI}$ under the $\mathcal{H}_0$ hypothesis is chi-square distributed with $2N$ degrees of freedom and is given by
\begin{equation}\label{eq:chi2pdfH0} 
\frac{2E_{CSI}}{{||\mathbf{w}_{CSI}^H\mathbf{h}[n]||}^2\sigma_{si}^2+||\mathbf{w}_{CSI}^H||^2\sigma_v^2}
\sim \chi_{2N}^2. 
\end{equation}
Based on $E_{CSI}$, the decision is taken as
\begin{equation}
E_{CSI} \underset{\mathcal{H}_0}{\overset{\mathcal{H}_1}{\gtrless}} \eta_{CSI},
\end{equation}
Using the statistics under $\mathcal{H}_0$ given in (\ref{eq:chi2pdfH0}), the threshold for Neyman-Pearson detector is calculated. In this case with a constraint on the false alarm probability $P_{fa}$
\cite{KayDetTheory} threshold is evaluated by
\begin{align}
&\eta_{CSI} \nonumber \\&\hspace{0.1cm}= \frac{{||\mathbf{w}_{CSI}^H\mathbf{h}[n]||}^2\sigma_{si}^2+||\mathbf{w}_{CSI}^H||^2\sigma_v^2}{2}\mathcal{Q}^{-1}_{\chi^2_{2N}}\left(P_{fa}\right).\label{eq:eta}
\end{align}

%%%%%%%%%%%%%%%%%%%%%%%%%%%%%%%%%
\section{BFED  Spectrum Sensing Scheme with Perturbed DoAs}\label{Sec:BFEF-PDoA}
%%%%%%%%%%%%%%%%%%%%%%%%%%%%%%%%%
In practical scenario, the DoAs are not known aprior. To estimate DoAs several methods such as  MUSIC algorithm, ESPRIT algorithm, etc\cite{Balanis_book1} are used in existing literature. In a multipath channel as different paths are  correlated these methods fail to identify DoAs\cite{spa_smoothing}. Thus, techniques like spatial smoothing are used to de-correlate the signals and then the above mentioned methods can be used to estimate the DoAs. In this paper, we abstract out the DoAs estimation schemes by adding an error or perturbation to the actual angles. Error is assumed to be zero mean Gaussian random variable and the error variance is derived using the CRLB. In this case, the CRLB is given by \cite{CRLB} 
\begin{equation}
\textrm{var}(\hat{\theta}) = \frac{6}{|\alpha_{il}|^2 M^3 N \SNR \sin^2{\theta}},
\end{equation}
where $\hat\theta$ = $\theta + \Delta\theta$ with $\theta$ referring to the true angle. Here the assumption is that $\hat\theta$ is unbiased estimator of $\theta$.
\begin{equation}
    \textrm{var}(\hat\theta) = \textrm{var}(\Delta\theta) \geq \textrm{CRLB}(\theta).
\end{equation}
Based on these estimates, the interference matrix is constructed using  the array steering vectors corresponding to all DoAs of interferer. It is given as
  \begin{equation}
      \mathbf{I}_i=[\mathbf{a}(\hat\theta_{i1}), \ \mathbf{a}(\hat\theta_{i2}),\ \ldots\ \mathbf{a}(\hat\theta_{iK}) ]^H,
  \end{equation}
 where $K$={$N_c$}{$N_r$}. Now, applying the SVD to obtain the null space, the matrix $\mathbf{I}_i$ can be rewritten as
\begin{equation}
      \mathbf{I}_i=\mathbf{U}_1\mathbf{\Sigma}_1[\mathbf{V}_{2}\mathbf{V}_{3}]^H,
\end{equation}
where $\mathbf{V}_{2}$ holds the first $K$ right singular vectors and  $\mathbf{V}_{3}$ holds the last $M-K$ right singular vectors. The square norm of projection matrix is given as
\begin{equation}
   ||{\mathbf{w}_{i}}||^2=\mathbf{w}^H\mathbf{V}_{3}{\mathbf{{V}_{3}}^H}\mathbf{w}.
\end{equation}
Further, beamforming gain is calculated as
\begin{equation}
B_{GF}=\mathbf{w}^H \mathbf{A} \mathbf{A}^H\mathbf{w},
\end{equation}
where $\mathbf{A}=[\mathbf{a}(\hat\theta_{p1}), \ \mathbf{a}(\hat\theta_{p2}),\ \ldots\ \mathbf{a}(\hat\theta_{pK}) ]$ is the matrix consisting of array steering vectors corresponding to the DoAs from PU. Similar to the full CSI case, here also we take the weighted sum of both the beamforming gain and the square norm of projection to formulate the multi-objective optimization problem as
\begin{equation}
\begin{split}
 \mathbf{W}_{3} &=  \argmax_{\mathbf{W}} \, \Big\{\textrm{Tr}\big((\lambda\, \mathbf{V}_{3} {\mathbf{V}_{3}}^H \\
& \qquad \ + 1-\lambda) \, \mathbf{A} \mathbf{A}^H \big) \mathbf{W}) \Big \},  \\
& \textrm{s.t.} 
 \quad [\mathbf{W}]_{kk}=\frac{1}{M}, 
 \quad \mathbf{W}\geq 0.    
 \end{split}
\end{equation}
Using $\mathbf{W}_3$, the analog beamforming weight  vector $\mathbf{w}_d$ is taken to be the scaled first column of the matrix $\mathbf{W_3}$ (given as $\mathbf{w_3}$), 
\begin{equation}
    {\bf{w}}_d =  \sqrt{M}{\bf{w}}_3.
\end{equation}
Now similar to \eqref{eq:mag} $\mathbf{w}_d$ is transformed to $\mathbf{w}_p$. The received signal is then multiplied by the weight vector given as
\begin{equation}
    y_p[n] = \mathbf{w}_{p}^H \mathbf{x}[n]
\end{equation}
while the energy $E_p$ is 
\begin{equation}
E_p = \sum_{n=1}^{N}|y_p[n]|^2.
\end{equation}
The decision is taken based on
\begin{equation}
E_p \underset{\mathcal{H}_0}{\overset{\mathcal{H}_1}{\gtrless}} \eta_p.
\end{equation}
\begin{comment}
Since $E_p$ is the sum of the squares of $N$ complex Gaussian random variables, $\frac{E_p}{\sigma_y^2/2}$ follow central chi-square $(\chi^2)$ distribution with $2N$ degrees of freedom under the hypotheses $\mathcal{H}_0$ \cite{KayDetTheory} so that 

\begin{equation} 
\frac{2E_p}{\sigma_{v}^2+\sigma_{si}^2}\sim \chi_{2N}^2\label{eq:chi2pdf H0}
\end{equation}
the threshold for this case can be obtained for a Neyman-Pearson detector with a constraint on the false alarm probability $P_{fa}$ \cite{KayDetTheory} by
\begin{equation}
\eta_p = \left(\frac{\sigma_{v}^2+\sigma_{si}^2}{2}\right)\, \mathcal{Q}^{-1}_{\chi^2_{2N}}\left(P_{fa}\right).\label{eq:eta1}
\end{equation}
\end{comment}
The energy $E_p$ is compared with the threshold $\eta_p$. The threshold is calculated empirically\footnote[2]{Threshold is calculated empirically by taking CDF of energy values under $\mathcal{H}_0$ and picking the energy value corresponding to the chosen $P_{fa}$ } under $\mathcal{H}_0$ because it is dependent on the CSI of the interfering channel.

%%%%%%%%%%%%%%%%%%%%%%%%%%%%%%%
\begin{comment}
\section{Simple Energy Detection }\label{Sec:ED}
%%%%%%%%%%%%%%%%%%%%%%%%%%%%%%%
In case of simple ED, which does not involve DoA estimation or beamforming, the received signal energy $E_s$ at $M$ antennas for $N$ time instances can be expressed as
\begin{eqnarray} \label{eq:SingleE}
E_s=\sum_{n=1}^{N}\sum_{m=1}^{M}|x_m[n]|^2.
\end{eqnarray}
The decision is taken based on
\begin{equation}
E_s \underset{\mathcal{H}_0}{\overset{\mathcal{H}_1}{\gtrless}} \eta_s.
\end{equation}
Since $E_s$ given in (\ref{eq:SingleE}) is the sum of the squares of $MN$ complex Gaussian random variables, $\frac{E_s}{\sigma_x^2/2}$ follow central chi-square $(\chi^2)$ distribution with $2MN$ degrees of freedom under the hypotheses $\mathcal{H}_0$ \cite{KayDetTheory} so that 
\begin{equation} 
\frac{2E_s}{\sigma_{v}^2+\sigma_{si}^2}\sim \chi_{2MN}^2\label{eq:chi2pdf H0}.
\end{equation}
Using Neyman-Pearson detector to detect the threshold in this case with a constraint on the false alarm probability $P_{fa}$
\cite{KayDetTheory} by
\begin{equation}
\eta_s = \left(\frac{\sigma_{v}^2+\sigma_{si}^2}{2}\right) \, \mathcal{Q}^{-1}_{\chi^2_{2MN}}(P_{fa}).\label{eq:eta1}
\end{equation}
%-------------------------
\end{comment}

%%%%%%%%%%%%%%%%%%%%%%%%%%%%
\section{Simulation Results}\label{Sec:Results}
%%%%%%%%%%%%%%%%%%%%%%%%%%%%
%\textcolor{red}{DoA=$40^o$, L=201??}

In this section, simulation results are provided to show the detection performance of various methods. The number of Monte Carlo realizations for estimating the probability of detection $(P_d)$ is considered as $1000$. Moreover, the various parameters are set as: the noise variance as $\sigma_v^2=1$, the sample size as $N=200$, and the number of antennas as $M=16$ unless specified otherwise. For all the methods, the probability of false alarm is set to $P_{fa}=0.1$, whereas the interference power $\sigma_{si}^2$ is set to 5 dB. Note that these values are chosen for simulation purposes. However, the analysis presented in this paper is valid for any arbitrary values of system parameters. The optimization problem to find the weight vector is solved in Matlab using the CVX package.

%----------------------------
\subsection{Optimal Value of $\lambda$}
%----------------------------
%-------------optimal lambda--------------%
\begin{figure}[t!]
	\centering
	\includegraphics[width=3.25in,height=2.4in]{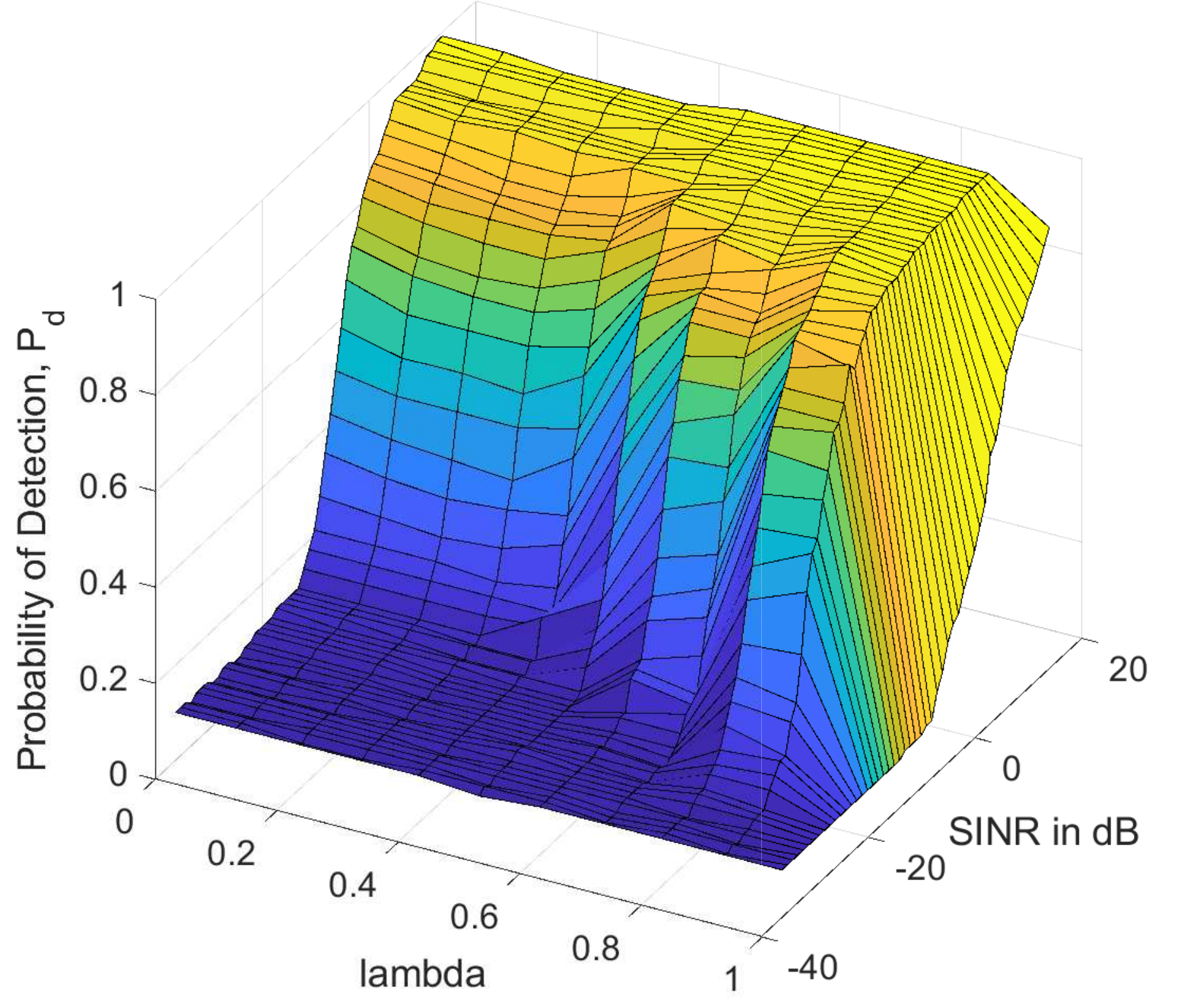}
	\vspace{-0.3cm}
    \caption{$P_d$ vs SINR (dB) plot for different values of $\lambda$.}
	\label{fig:pd_snr_M4}
	\vspace{-0.5cm}
\end{figure}
%-----------------------------------------%
%--------------------%
To obtain the value of $\lambda$, which maximizes the probability of detection for the proposed BFED (with estimated DoAs) in the considered scenario, the algorithm was run for varying SINR as well as $\lambda$ in steps of 0.1 from 0 to 1. Fig. 2 shows the detection performance as a function of SINR (in dB) and $\lambda$. It can be seen that for a given $\lambda$, the detection probability increases with increasing SINR as expected. Also, for a given SINR, the detection probability increases as $\lambda$ increases with an exception at $\lambda=1$. At $\lambda=1$, the objective function completely favors beamforming gain ignoring interference, which results in deterioration of detection performance. For a given SINR, $P_d$ is maximized at $\lambda = 0.9$ for the considered scenario. This is expected as the interference power is higher than the PU signal power at low SNRs.  Thus, in the rest of the simulations, $\lambda$ is considered as $0.9$.
%----------------------------
\subsection{$P_d$ vs SINR and ROC curves}
%----------------------------
\begin{figure}[t]
	\centering
	\includegraphics[width=3.0in,height=2.3in]{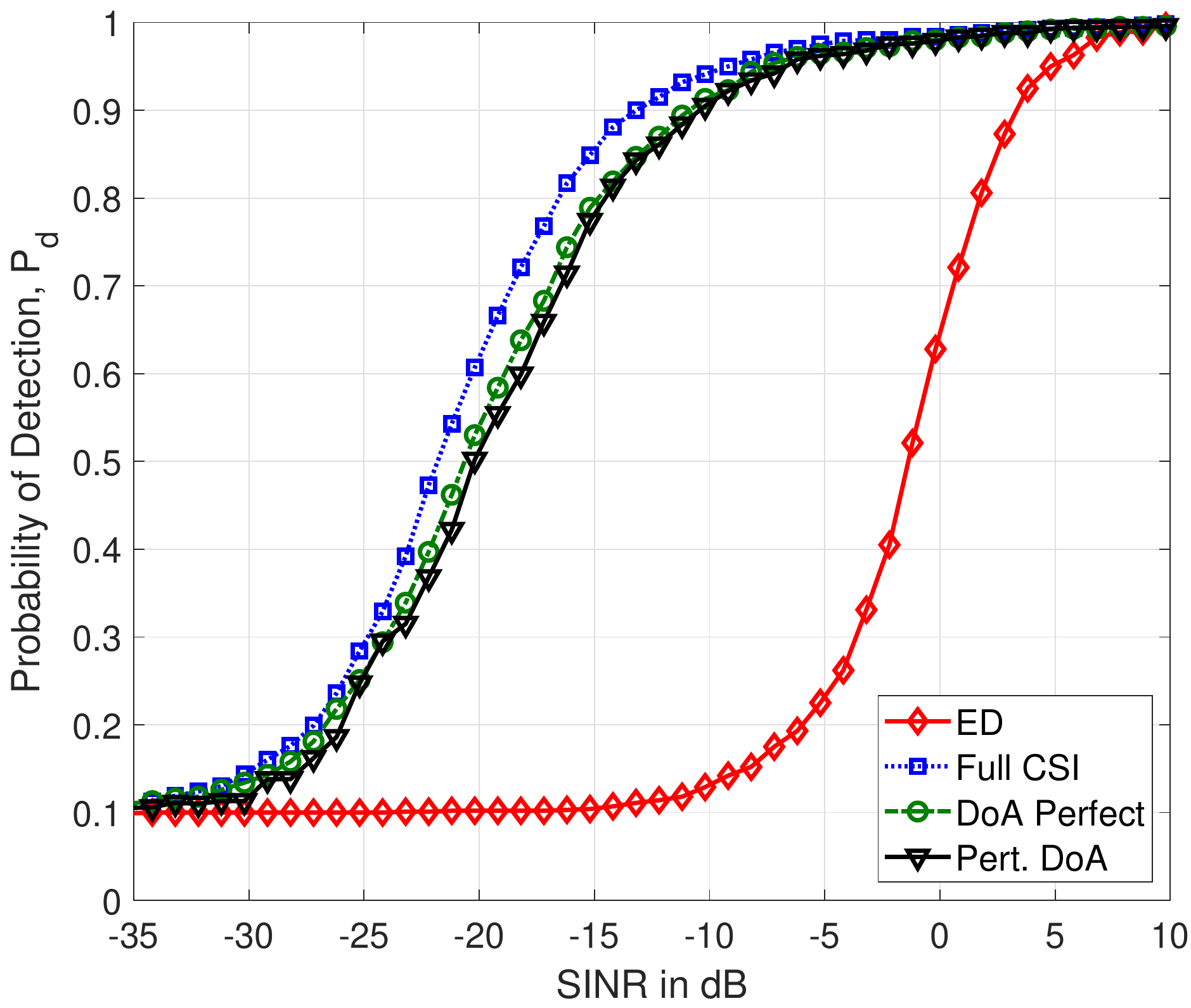}
	\vspace{-0.3cm}
    \caption{$P_d$ vs SINR (dB) plot for $N=200$ for simple ED and the proposed BFED algorithms with different degrees of CSI information.}
    \label{fig:pd_snr_M4}
\end{figure} 
%--------------------%
%--------------------%
\begin{figure}
	\centering
\includegraphics[width=3.0in,height=2.3in]{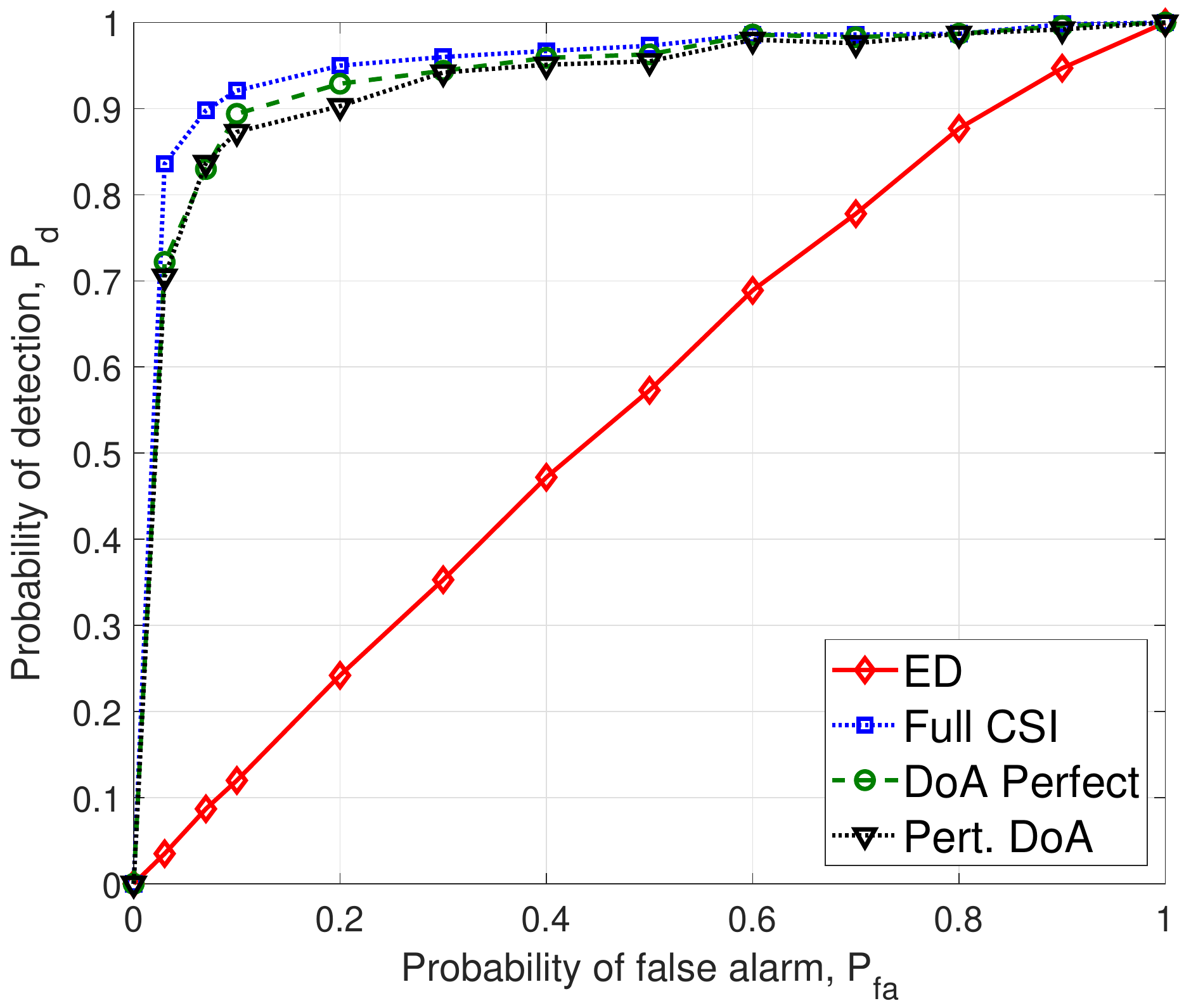}
\vspace{-0.3cm}
    \caption{$P_d$ vs $P_{fa}$ plots for SINR = -5 (dB).}
    	\vspace{-0.5cm}
	\label{fig:ROC}
\end{figure} 
%---
Fig. \ref{fig:pd_snr_M4} shows the probability of detection $(P_d)$ vs SINR plots with $M=16$ , $N_c = 2$ and $N_r = 2$ for four methods: (i) simple energy detection, (ii) BFED with full CSI, (iii) BFED with perfect DoAs (iv) BFED with perturbed DoAs. In case of simple ED, the received signal energy $E_s$ is compared with a threshold $\eta_s = \left(\frac{\sigma_{v}^2+\sigma_{si}^2}{2}\right) \, \mathcal{Q}^{-1}_{\chi^2_{2MN}}(P_{fa}).\label{eq:eta1}$ using Neyman-Pearson detector \cite{KayDetTheory}. BFED with perfect DoAs serves as an upper bound for BFED with perturbed DoAs. The first observation is that the BFED schemes give significant gain as compared to simple energy detection. The second observation is that BFED with full CSI (both DoAs and fading coefficients completely known) performs the best among the compared schemes. The third observation is that the performance degradation going from full CSI to only knowing or estimating DoAs is minor even though the weight vector is chosen from a restricted set of vectors in which the magnitude of each element is constant. The fourth observation is that there is no significant difference in performance between DoAs perfectly known and perturbed DoAs cases for the considered scenario. The similar observations can also be seen in ROC curves presented in Fig. \ref{fig:ROC}.
%----------------%
%--------------------%
\begin{figure}[t!]
\centering
\includegraphics[width=3.0in,height=2.3in]{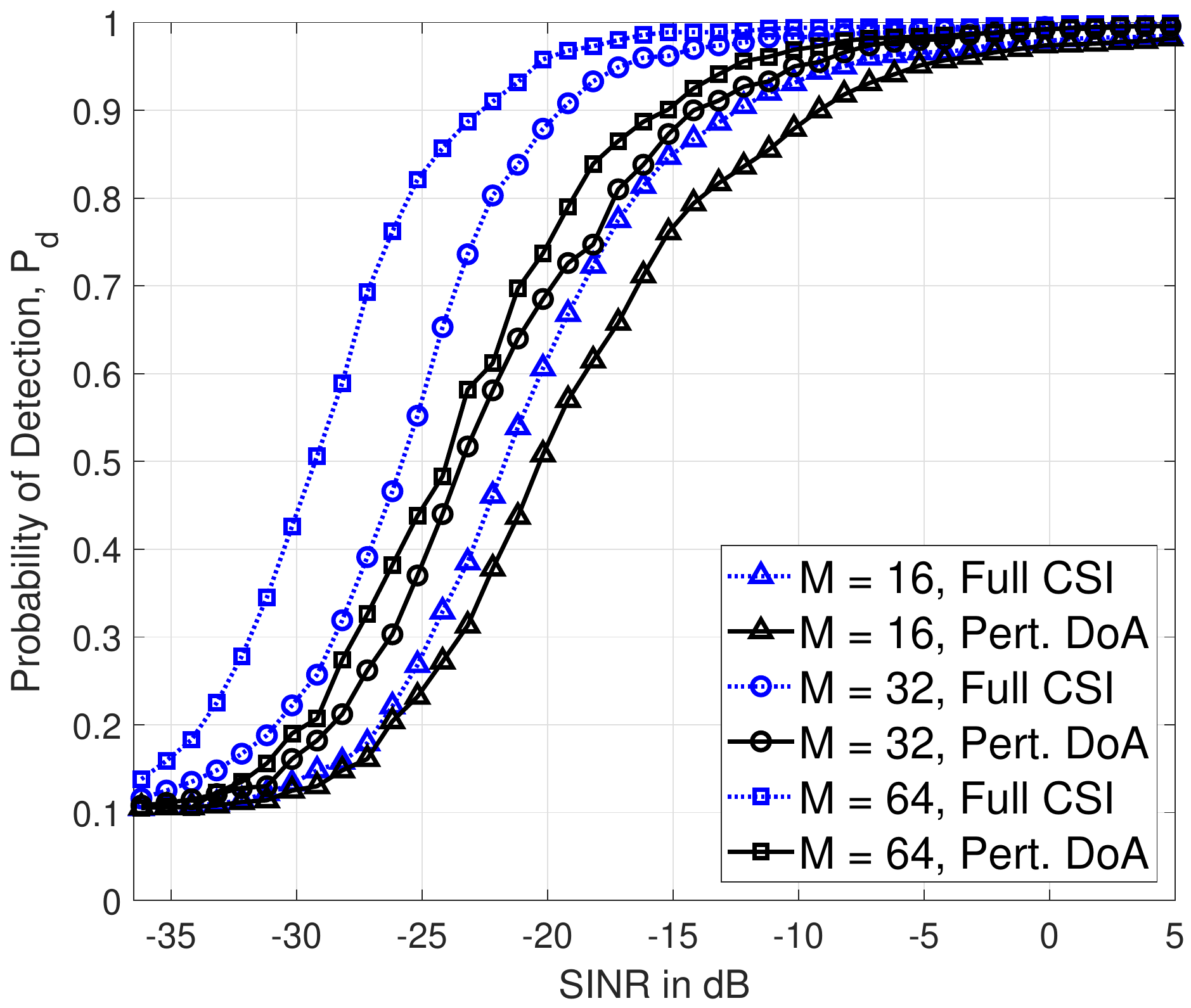}
	\vspace{-0.3cm}
\caption{$P_d$ vs SINR (dB) plots for BFED with perturbed DoAs and perfect CSI for different number of receiver antennas.}
	\vspace{-0.2cm}
\label{fig:pd_snr_Pf}
\end{figure}

%--------------------%
Fig. \ref{fig:pd_snr_Pf} shows the detection performances of BFED schemes with full CSI and perturbed DoAs for the different number of receiver antennas. The first observation is that as the number of antennas is increased, there is a significant improvement in the detection performance. The second observation is that the gap in between the performances of BFED schemes with full CSI and with perturbed DoAs increases with an increase in the number of antennas. This is because the beamwidth decreases with the increase in antennas, and a small error in DoA results in a significant performance decrease. 
\begin{comment}
\textcolor{red}{Does not make sense!!}. 
\end{comment}
%--------------------%
\begin{figure}[t!]
	\centering
\includegraphics[width=3.0in,height=2.3in]{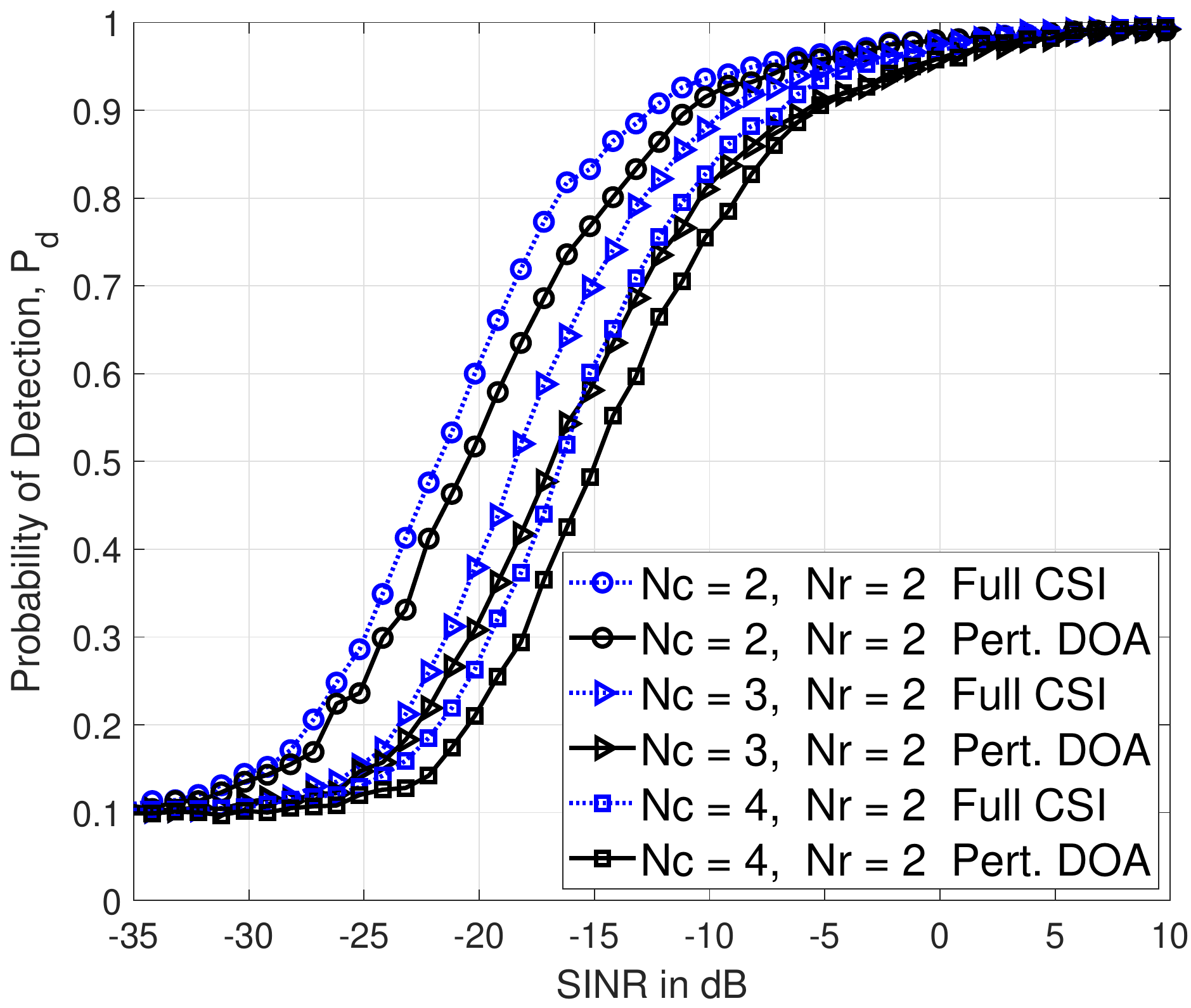}
    	\vspace{-0.3cm}
\caption{$P_d$ vs SINR (dB) plots for varying numbers of clusters and rays for the proposed BFED schemes with perturbed DoAs and full CSI.}
	\label{fig:pd_snr_N_M}
	\vspace{-0.5cm}
\end{figure} 
%---

Fig. \ref{fig:pd_snr_N_M} shows the performance comparison between the BFED schemes with perturbed DoAs and full CSI for different values of clusters and rays. It can be observed from the figure, as the number of clusters increases, the detection performance degrades in general. This happens because the power is now split into many rays and clusters, and with the constant magnitude constraint on the weight
vector, the received power decreases that in turn decreases the detection probability. 
%-------------------------
\section{Conclusion}\label{Sec:Conclusion}
%--------------------------
%In this paper, BFED based spectrum sensing scheme is proposed. In this paper, BFED based spectrum sensing has been proposed for the scenarios when either full CSI is available or DoAs are estimated (perturbed DoAs). The performance of this method has been compared with that of full CSI. In the Full CSI case, the receiver has the overhead of estimating both angle and fading coefficients. In the proposed method, beamforming is done using only DoA information and requires only one RF chain, by which the computational complexity is reduced significantly.
In this paper, two BFED schemes have been proposed to detect a PU in the presence of an interferer in a cognitive mmWave network. The first beamforming scheme is designed under the assumption of full CSI to demonstrate the bound on the system performance. The second scheme is designed with no CSI assumption and works with perturbed DoAs. In both cases, the beamforming weights are found by solving the  optimization problem for maximizing beamforming gain towards the intended PU and reducing interference from the interferer. The distribution of the test statistics has been derived under the null hypothesis, and the threshold of a Neyman-Pearson detector found analytically for full CSI case. However, the threshold for perturbed DoAs case was obtained empirically since the threshold under null hypothesis is dependent on the CSI of the interfering channel. The proposed schemes have shown significant improvement in the detection performance as compared to the traditional ED. Also, the performance degradation from full CSI to perturbed DoAs scheme is minimal for the considered scenario. In future, we plan to extend this work using real measurements based mmWave channel model along with multiple antennas at the PU and interferer.
\begin{comment}
{

\section*{Acknowledgement}
 The research work of M. Madhuri Latha is partially supported by the Visvesvaraya Ph.D. scheme for Electronics and Information technology (IT), Ministry of Electronics and IT, Government of India, and G. Narayanamma Institute of Technology and Science (for Women), Hyderabad, India.
 }
\end{comment}
\bibliographystyle{IEEEtran}
\bibliography{IEEEabrv,reference.bib}{}	

\begin{thebibliography}{10}
\providecommand{\url}[1]{#1}
\csname url@samestyle\endcsname
\providecommand{\newblock}{\relax}
\providecommand{\bibinfo}[2]{#2}
\providecommand{\BIBentrySTDinterwordspacing}{\spaceskip=0pt\relax}
\providecommand{\BIBentryALTinterwordstretchfactor}{4}
\providecommand{\BIBentryALTinterwordspacing}{\spaceskip=\fontdimen2\font plus
\BIBentryALTinterwordstretchfactor\fontdimen3\font minus
  \fontdimen4\font\relax}
\providecommand{\BIBforeignlanguage}[2]{{%
\expandafter\ifx\csname l@#1\endcsname\relax
\typeout{** WARNING: IEEEtran.bst: No hyphenation pattern has been}%
\typeout{** loaded for the language `#1'. Using the pattern for}%
\typeout{** the default language instead.}%
\else
\language=\csname l@#1\endcsname
\fi
#2}}
\providecommand{\BIBdecl}{\relax}
\BIBdecl

\bibitem{Niu2015}
Y.~Niu, Y.~Li, D.~Jin, L.~Su, and A.~V. Vasilakos, ``A survey of millimeter
  wave communications (mmwave) for {5G}: opportunities and challenges,''
  \emph{Wireless Networks}, vol.~21, no.~8, pp. 2657--2676, Nov 2015.

\bibitem{Kutty2016}
S.~{Kutty} and D.~{Sen}, ``Beamforming for millimeter wave communications: An
  inclusive survey,'' \emph{IEEE Commun. Surveys and Tutorials}, vol.~18,
  no.~2, pp. 949--973, 2016.

\bibitem{Hosseini2016}
H.~{Hosseini}, A.~{Anpalagan}, K.~{Raahemifar}, S.~{Erkucuk}, and S.~{Habib},
  ``Joint wavelet-based spectrum sensing and {FBMC} modulation for cognitive
  mmwave small cell networks,'' \emph{IET Commun.}, vol.~10, no.~14, pp.
  1803--1809, 2016.

\bibitem{Gupta2016}
A.~K. {Gupta}, J.~G. {Andrews}, and R.~W. {Heath}, ``On the feasibility of
  sharing spectrum licenses in mmwave cellular systems,'' \emph{IEEE Trans.
  Commun.}, vol.~64, no.~9, pp. 3981--3995, 2016.

\bibitem{Tsinos2020}
C.~G. {Tsinos}, S.~{Chatzinotas}, and B.~{Ottersten}, ``Hybrid analog-digital
  transceiver designs for multi-user {MIMO} mmwave cognitive radio systems,''
  \emph{IEEE Trans. Cognitive Commun. Networking}, vol.~6, no.~1, pp. 310--324,
  2020.

\bibitem{Park2018}
J.~{Park}, J.~G. {Andrews}, and R.~W. {Heath}, ``Inter-operator base station
  coordination in spectrum-shared millimeter wave cellular networks,''
  \emph{IEEE Trans. Cognitive Commun. Networking}, vol.~4, no.~3, pp. 513--528,
  2018.

\bibitem{mmWaveCR}
Y.~{Song}, W.~{Yang}, X.~{Yang}, Z.~{Xiang}, and B.~{Wang}, ``Physical layer
  security in cognitive millimeter wave networks,'' \emph{IEEE Access}, vol.~7,
  pp. 109\,162--109\,180, 2019.

\bibitem{Thesis}
S.~Chaudhari, ``Spectrum sensing for cognitive radios: Algorithms, performance,
  and limitations,'' Ph.D. dissertation, Aalto University, Finland, Nov. 2012.

\bibitem{NCC}
M.~M. {Latha}, P.~B. {Gohain}, and S.~{Chaudhari}, ``Low complexity two-stage
  sensing using energy detection and beamforming,'' in \emph{Twenty Fourth
  National Conf. Commun. (NCC)}, 2018, pp. 1--6.

\bibitem{BF_Sensing_1}
K.~Bouallegue, I.~Dayoub, M.~Gharbi, and K.~Hassan, ``A cost-effective approach
  for spectrum sensing using beamforming,'' \emph{Physical Commun.}, vol.~22,
  pp. 1--8, 2017.

\bibitem{BF_Sensing_2}
D.~Wilcox, E.~Tsakalaki, A.~Kortun, T.~Ratnarajah, C.~B. Papadias, and
  M.~Sellathurai, ``On spatial domain cognitive radio using single-radio
  parasitic antenna arrays,'' \emph{IEEE J. Select. Areas Commun.}, vol.~31,
  no.~3, pp. 571--580, March 2013.

\bibitem{Makarfi2012}
A.~U. {Makarfi} and K.~A. {Hamdi}, ``Efficiency of energy detection for
  spectrum sensing in the presence of non-cooperating secondary users,'' in
  \emph{IEEE Global Commun. Conf. (GLOBECOM)}, 2012, pp. 4939--4944.

\bibitem{Lagunas2014}
E.~{Lagunas} and M.~{Nájar}, ``Compressed spectrum sensing in the presence of
  interference: Comparison of sparse recovery strategies,'' in \emph{22nd
  European Signal Processing Conf. (EUSIPCO)}, 2014, pp. 1721--1725.

\bibitem{ANTS}
M.~M. Latha, S.~K.~C. Dara, and S.~Chaudhari, ``Beamformed sensing using
  dominant doa in cognitive mmwave network,'' in \emph{2020 IEEE International
  Conference on Advanced Networks and Telecommunications Systems (ANTS)}, 2020,
  pp. 1--6.

\bibitem{spectrum_sensing_survey}
F.~Awin, E.~Abdel-Raheem, and K.~Tepe, ``Blind spectrum sensing approaches for
  interweaved cognitive radio system: A tutorial and short course,'' \emph{IEEE
  Communications Surveys Tutorials}, vol.~21, no.~1, pp. 238--259, 2019.

\bibitem{channel}
O.~E. {Ayach}, S.~{Rajagopal}, S.~{Abu-Surra}, Z.~{Pi}, and R.~W. {Heath},
  ``Spatially sparse precoding in millimeter wave {MIMO} systems,'' \emph{IEEE
  Trans. Wireless Commun.}, vol.~13, no.~3, pp. 1499--1513, 2014.

\bibitem{CRLB}
D.~{Fan}, F.~{Gao}, Y.~{Liu}, Y.~{Deng}, G.~{Wang}, Z.~{Zhong}, and
  A.~{Nallanathan}, ``Angle domain channel estimation in hybrid millimeter wave
  massive {MIMO} systems,'' \emph{IEEE Trans. Wireless Commun.}, vol.~17,
  no.~12, pp. 8165--8179, 2018.

\bibitem{Jiang2019}
L.~{Jiang} and H.~{Jafarkhani}, ``Multi-user analog beamforming in millimeter
  wave {MIMO} systems based on path angle information,'' \emph{IEEE Trans.
  Wireless Commun.}, vol.~18, no.~1, pp. 608--619, 2019.

\bibitem{KayDetTheory}
S.~Kay, \emph{Fundamental of Statistical Signal Processing: Volume II Detection
  Theory}.\hskip 1em plus 0.5em minus 0.4em\relax Prentice Hall, 1998.

\bibitem{Balanis_book1}
C.~A. Balanis, \emph{Introduction to Smart Antennas}.\hskip 1em plus 0.5em
  minus 0.4em\relax Morgan and Claypool, 2007.

\bibitem{spa_smoothing}
S.~U. {Pillai} and B.~H. {Kwon}, ``Forward/backward spatial smoothing
  techniques for coherent signal identification,'' \emph{IEEE Trans. Acoust.,
  Speech, Signal Pro}, vol.~37, no.~1, pp. 8--15, 1989.

\end{thebibliography}

\end{document}